\documentstyle[11pt,twoside,epsf]{article}

\begin{document}

\title{Large-scale Simulation of the Two-dimensional Kinetic Ising Model}

\author{Andreas Linke\thanks{\tt linke@tphys.uni-heidelberg.de},
				Dieter W. Heermann\\
Institut f\"ur theoretische Physik \\
Universit\"at Heidelberg \\
Philosophenweg 19 \\
D-69120 Heidelberg \\
Germany \\
and\\
Interdisziplin\"ares Zentrum\\
f\"ur wissenschaftliches Rechnen\\
der Universit\"at Heidelberg\\ {} \\
Peter Altevogt\\
Heidelberg Scientific \& Technical Center\\
IBM Deutschland Informationssysteme GmbH\\
D-69115 Heidelberg\\ {} \\
        Markus Siegert\\
Regionales Rechenzentrum\\
der Universit\"at zu K\"oln\\
Robert--Koch--Str. 10\\
D-50931 K\"oln\\
  }

  \vspace {3ex}

\maketitle

\begin{abstract}
\noindent We present Monte Carlo simulation results for the dynamical
critical exponent $z$ of the two-dimensional kinetic Ising model using
a lattice of size $10^6 \times 10^6$ spins.
We used Glauber as well as Metropolis
dynamics. The $z$-value of $2.16\pm 0.005$
was calculated from the magnetization and energy
relaxation from an ordered state towards the equilibrium state at $T_c$.

\vspace{1cm}
{\em submitted to Physica A}
\end{abstract}

\vfill\eject

The precise numerical value of the dynamical critical
exponent $z$, which characterizes the critical slowing down at a
second-order phase-transition~\cite{trans}, has not been conclusively
calculated.
This note presents an effort
to calculate the value of $z$ using a system of unprecedented size.

In simulations specific attention has been given to calculate $z$
for the Ising model because of its simplicity and its model character as
a representant for a universality class. The Ising model is defined by
the Hamiltonian

\begin{eqnarray}
{\cal H}_{\mbox{\scriptsize Ising}}(s) = - J \sum_{<ij>}s_is_j\qquad , \qquad
s_{i} = \pm 1
\end{eqnarray}

\noindent where $\left<ij\right>$
are nearest-neighbour pairs of lattice sites. The
exchange coupling $J$ is restricted to be positive (ferro-magnetic).

The dynamics of the system is specified by the transition probability of
the Markov chain which will be realized by a Monte Carlo
algorithm~\cite{Heermann1,Kalos,Binder-Heermann}. We used two transition
probabilities

\begin{itemize}
  \item {\bf Metropolis}
            $P(s_i \rightarrow s_i') = \min \{1,\exp\{-\Delta E )\}$

  \item {\bf Glauber}
            $P(s_i \rightarrow s_i') = \frac{1}{2}\{ 1 - s_i\tanh (E_i/k_bT)\},
            \qquad E_i = J \sum_{<i,j>}s_j$
\end{itemize}

\noindent Time $t$ in this context is measured in Monte Carlo steps
per spin. One
Monte Carlo step (MCS) per lattice site, i.e.\ one sweep through the
entire lattice comprises one time unit.
Neither magnetization nor energy
are preserved in the model.

Our system size, on which the analysis will
be based on, was $10^6 \times 10^6$.
A new algorithm~\cite{Linke-Heermann-Altevogt} was employed
to simulate this very large lattice.
Basically, instead of storing one lattice configuration and iterating in time
we store only part of the lattice configuration at all time steps and iterate
through the lattice. Thus we were able to reduce the memory requirements
from approximately 100 GB to 20 MB.

It is generally believed that one can calculate $z$ from the relaxation
into equilibrium. In the theory of dynamical critical phenomena
\cite{Hohenberg1977}, critical slowing down is expressed as
the divergence of the linear relaxation time $\tau_M^{(l)}$ of the order
parameter $M$ (the magnetization) as one approaches criticality ($\theta
\rightarrow 0^+$ with $\theta = \left(T-T_c\right)/T_c$)

\begin{equation}
\tau_M^{(l)} \sim \theta^{-\nu z},
\label{eq:z}
\end{equation}

\noindent where $\sim$ stands for asymptotic proportionality and
$\nu$ is the static critical exponent of the correlation
length $\xi$. The exponent $z$ is called  the dynamical critical exponent.
In analogy with finite-size scaling in the theory of
static critical phenomena one may
infer the following finite-size scaling relation for the linear
relaxation time

\begin{equation}
\tau_M^{(l)} \sim L^z \tilde{f}(\theta L^{\frac{1}{\nu}}),
\label{eq:FSSz}
\end{equation}

\noindent where $\tilde{f}(z^\prime)$ is a scaling function regular at
$z^\prime=0$.  Relation (\ref{eq:FSSz}) is valid for $L \gg 1$ and
$\theta \ll 1$.

\noindent A value for $z$ can be obtained for very large
system sizes  from the relaxation of the magnetization into equilibrium from
another equilibrium state~\cite{Ito}.
If the system size tends to infinity and we are at criticality,
the above scaling relaxation implies

\begin{equation}
M(t) \sim t^{-\beta/\nu z} \quad .
\label{eq:m}
\end{equation}
For the two-dimensional Ising model, the static critical exponents
$\beta=1/8$ and $\nu = 1$ as well as the critical temperature
$T_c=1/(\frac{1}{2}\log(1+\sqrt{2}))$ are exactly known.

Two independent calculations were carried out by groups from Heidelberg
using Metropolis dynamics and checkerboard updates
and Cologne
using Glauber dynamics and typewriter update.
\vspace{1cm}

\noindent {\bf Results from the Heidelberg group} \hspace {0.5cm}
During the course of our simulation we have monitored the magnetization
$M = (1/L^2)\sum s_i$ as a function of the number of sweeps $t$
(MCS) through the lattice.
We averaged over two independent configurations
(cf.\ implementation of the
algorithm in~\cite{Linke-Heermann-Altevogt}).
Table~\ref{data_linke} shows the raw data.

\begin{figure}
\epsfxsize=12cm
\leavevmode
\centering
\epsffile{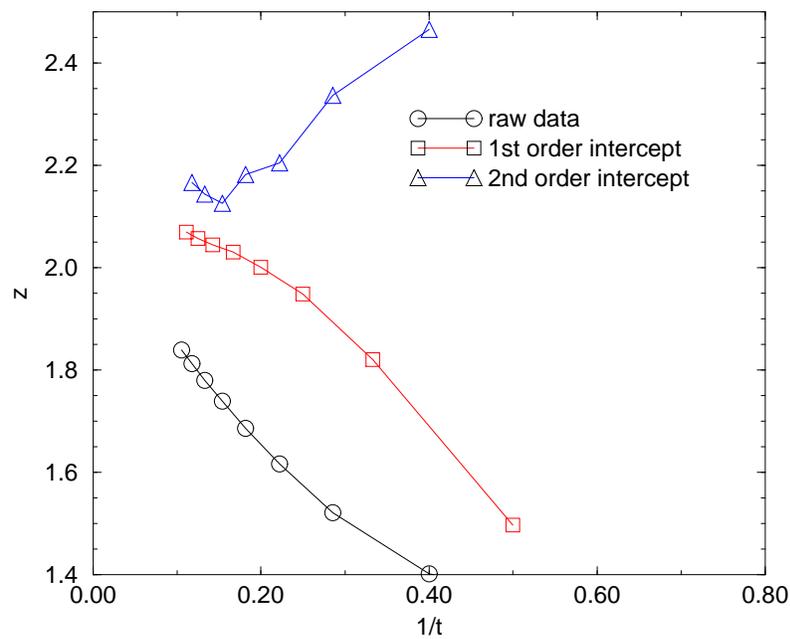}
 \caption{  \em Plot of the critical exponent $z$ vs.\ $1/t$
(Heidelberg group).}
 \label{z_linke}
\end{figure}

To determine the value of $z$ from our data
we calculated the slope in a log-log plot using
\[
\frac{1}{z_i} = -8 \frac{\log(m_i)-\log(m_{i-1})}
{\log(i)-\log(i-1)}
\]
and grouped data points
for a linear least-square fit. Plotted in figure~\ref{z_linke} is the
slope
and the first and second order
intercepts of the linear fits to these slopes for $1/t\rightarrow 0$
against the inverse time.
We obtain a value of $z\approx 2.15\pm 0.02$. This value is in agreement with
results from recent series expansion calculations \cite{Adler,Dammann},
damage spreading simulations \cite{Grassberger} as well as other
large scale simulates \cite{Munkel,Stauffer94}.
\vspace{1cm}

\noindent {\bf Results from the Cologne group} \hspace {0.5cm}
Shocked by the huge lattices from the Heidelberg group, we slightly
modified and adapted the step method presented in
\cite{Linke-Heermann-Altevogt} to typewriter update, which was used in
our previous calculations with full lattice storage
\cite{Siegert-Stauffer}. The use of typewriter update instead of
checkerboard update used by the Heidelberg group, simplified this
method considerable.  Thereby and with optimization techniques similar
to multi-spin coding our program reached 3.8 MUpdate/s on an IBM
RS6000/990 (Power/2). With this program we calculated 50 iterations of
a $10^6 \times 10^6$ lattice on an 8 processor IBM SP1 (Glauber
dynamic, user time 31 days). Beside the magnetization the energy was
obtained as well. Table~\ref{data_siegert} shows the data.

Similar to the magnetization one can also calculate the exponent $z$
from the relaxation of the energy \cite{Stauffer94}:

\begin{equation}
E(t) - E_{\mbox{\scriptsize critical}}
\sim t^{-(1-\alpha)/\nu z} = t^{-1/z} \quad {\rm (in~two~dimensions)}
\end{equation}

\begin{figure}
\epsfxsize=13cm
\leavevmode
\centering
\epsffile{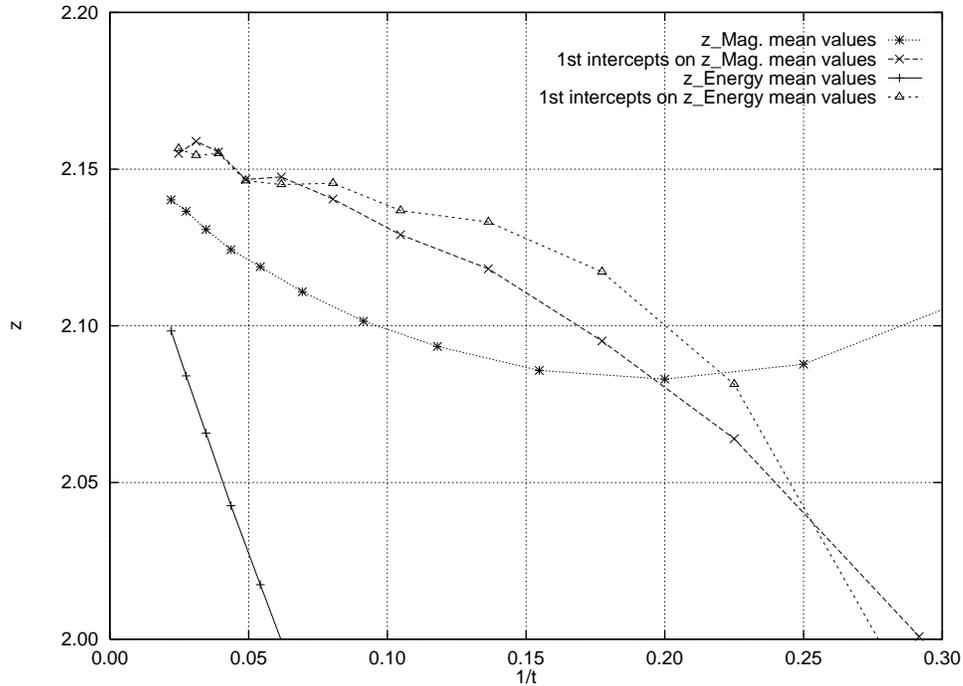}
\caption{  \em Plot of the critical exponent $z$ vs.\ $1/t$ (Cologne group).}
\label{z_siegert}
\end{figure}

For the determination of the value of $z$ we calculated an arithmetic
mean value of some data points. The number of gathered data points is
calculated by $\lceil {\rm iteration} / 5 \rceil$
Figure~\ref{z_siegert} shows this and the first order intercepts. The
magnetization data $z_{\mbox{\scriptsize Mag}}$ as well as the energy data
$z_{\mbox{\scriptsize Energy}}$
lead to a z-value of 2.16 $\pm$ 0.005 for $1/t \rightarrow 0$.

\begin{table}
\scriptsize\centering
\begin{tabular}{l|ll}
T &$m_1$ (Metropolis)&$m_2$ (Metropolis) \\
\hline
0 & 1.00000000  & 1.00000000 \\
1 & 0.92097430  & 0.92097417 \\
2 & 0.86321233  & 0.86321220 \\
3 & 0.83255023  & 0.83255007 \\
4 & 0.81309636  & 0.81309770 \\
5 & 0.79918281  & 0.79918239 \\
6 & 0.78845203  & 0.78845225 \\
7 & 0.77976347  & 0.77976351 \\
8 & 0.77248408  & 0.77248426 \\
9 & 0.76623362  & 0.76623401 \\
10 & 0.76076646  & 0.76076750 \\
\end{tabular}
\caption{\em Raw data for the calculation from the Heidelberg group.
\label{data_linke}}
\end{table}

\vspace{1cm}

\begin{table}
\scriptsize\centering
\begin{tabular}{r|r|r||r|r|r||r|r|r}
  T & m (Glauber)& e (Glauber)&  T & m (Glauber)& e (Glauber) &
	T & m (Glauber)& e (Glauber) \\
\hline
  0 & 1.00000000 & 0.00000000 & 17 & 0.78287989 & 0.49666620 &
	34 & 0.75161947 & 0.52225099 \\
  1 & 0.92387181 & 0.07612802 & 18 & 0.78024367 & 0.49915815 &
	35 & 0.75034603 & 0.52312965 \\
  2 & 0.88930745 & 0.27878672 & 19 & 0.77775922 & 0.50144418 &
	36 & 0.74910920 & 0.52396988 \\
  3 & 0.86826671 & 0.35298481 & 20 & 0.77541236 & 0.50355210 &
	37 & 0.74790926 & 0.52477812 \\
  4 & 0.85343942 & 0.39143993 & 21 & 0.77318752 & 0.50550180 &
	38 & 0.74674372 & 0.52555236 \\
  5 & 0.84208738 & 0.41546570 & 22 & 0.77107242 & 0.50731384 &
	39 & 0.74561180 & 0.52629785 \\
  6 & 0.83292949 & 0.43221206 & 23 & 0.76905750 & 0.50900323 &
	40 & 0.74450920 & 0.52701310 \\
  7 & 0.82527573 & 0.44472865 & 24 & 0.76713498 & 0.51058358 &
	41 & 0.74343671 & 0.52770277 \\
  8 & 0.81871664 & 0.45453473 & 25 & 0.76529694 & 0.51206432 &
	42 & 0.74238953 & 0.52836585 \\
  9 & 0.81298342 & 0.46248268 & 26 & 0.76353643 & 0.51345741 &
	43 & 0.74136891 & 0.52900757 \\
 10 & 0.80789697 & 0.46910024 & 27 & 0.76184711 & 0.51476957 &
 44 & 0.74037224 & 0.52962886 \\
 11 & 0.80333031 & 0.47472005 & 28 & 0.76022277 & 0.51601052 &
 45 & 0.73940143 & 0.53022785 \\
 12 & 0.79918884 & 0.47957117 & 29 & 0.75865931 & 0.51718485 &
 46 & 0.73845357 & 0.53080592 \\
 13 & 0.79540222 & 0.48381554 & 30 & 0.75715202 & 0.51830041 &
 47 & 0.73752722 & 0.53136430 \\
 14 & 0.79191812 & 0.48756875 & 31 & 0.75569749 & 0.51936114 &
 48 & 0.73662030 & 0.53190743 \\
 15 & 0.78869300 & 0.49091859 & 32 & 0.75429320 & 0.52036977 &
 49 & 0.73573540 & 0.53243573 \\
 16 & 0.78568804 & 0.49393126 & 33 & 0.75293487 & 0.52133128 &
 50 & 0.73486859 & 0.53294310 \\
\end{tabular}
\caption{\em Raw data for the calculation by the Cologne group.}
\label{data_siegert}
\end{table}

\vspace{1cm}
\noindent {\large\bf Acknowledgment} \hspace {0.5cm} Partial support
from the BMFT
project 0326657D and EU project CPACT~930105 (PL296476) is gratefully
acknowledged.
Part of this work was funded by a Stipendium of the Graduiertenkolleg
``Modellierung und Wissenschaftliches Rechnen in Mathematik und
Natur\-wis\-sen\-schaf\-ten'' at the IWR Heidelberg.

We thank D.\ Stauffer for numerous fruitful
discussions and scientific advice.
We thank the Leibniz Supercomputing Center
Munich
for the generous
grant of computation time at the local IBM 9076 SP2 parallel computer
and the
{\em Zentrum f\"ur paralleles Rechnen} Cologne
for the possibility to use their IBM SP1.

\vfill\eject
\end{document}